\documentclass[secnumarabic, graphics,floatfix, nofootinbib,tightenlines,nobibnotes, aps, prl, 12pt]{revtex4-2}
\usepackage{graphicx}
\usepackage[english]{babel}
\usepackage[utf8]{inputenc}
\usepackage{amsmath,amssymb}
\usepackage{mathrsfs}
\usepackage{amsfonts}
\usepackage{multirow}
\usepackage{pstricks}
\usepackage{float}
\usepackage{subfigure}
\usepackage{color}
\usepackage{epsfig}
\usepackage{pst-tools}

\begin{document}
 \newcommand{\bq}{\begin{equation}}
 \newcommand{\eq}{\end{equation}}
 \newcommand{\bqn}{\begin{eqnarray}}
 \newcommand{\eqn}{\end{eqnarray}}
 \newcommand{\nb}{\nonumber}
 \newcommand{\lb}{\label}
 
\title{Asymptotical quasinormal mode spectrum for piecewise approximate effective potential}

\author{Wei-Liang Qian$^{1,2,3}$}\email[E-mail: ]{wlqian@usp.br (corresponding author)}
\author{Kai Lin $^{4, 3}$}
\author{Cai-Ying Shao$^{5}$}
\author{Bin Wang$^{1}$}
\author{Rui-Hong Yue$^{1}$}

\affiliation{$^{1}$ Center for Gravitation and Cosmology, College of Physical Science and Technology, Yangzhou University, 225009, Yangzhou, China}
\affiliation{$^{2}$ Institute for theoretical physics and cosmology, Zhejiang University of Technology, 310032, Hangzhou, China}
\affiliation{$^{3}$ Escola de Engenharia de Lorena, Universidade de S\~ao Paulo, 12602-810, Lorena, SP, Brazil}
\affiliation{$^{4}$ Hubei Subsurface Multi-scale Imaging Key Laboratory, Institute of Geophysics and Geomatics, China University of Geosciences, 430074, Wuhan, Hubei, China}
\affiliation{$^{5}$ MOE Key Laboratory of Fundamental Physical Quantities Measurement, Hubei Key Laboratory of Gravitation and Quantum Physics, PGMF, and School of Physics, Huazhong University of Science and Technology, 430074, Wuhan, Hubei, China}

\date{Jan. 30th, 2024}

\begin{abstract}

It was pointed out that the black hole quasinormal modes resulting from a piecewise approximate potential are drastically distinct from those pertaining to the original black hole metric.
In particular, instead of lining up parallel to the imaginary axis, the spectrum is found to stretch out along the real axis.
In this work, we prove that if there is a single discontinuity in the effective potential, no matter how insignificant it is, the asymptotical behavior of the quasinormal modes will be appreciably modified.
Besides showing numerical evidence, we give analytical derivations to support the above assertion even when the discontinuity is located significantly further away from the maximum of the potential and/or the size of the step is arbitrarily small.
Moreover, we discuss the astrophysical significance of the potential implications in terms of the present findings.

\end{abstract}

\maketitle

\section{I. Introduction}

Reminiscent of the prominent role carried by the normal modes in a conserved physical system, crucial information about an open system is encoded in terms of its counterpart, known as quasinormal modes.
In the context of black hole configurations, the quasinormal modes describe the dissipative temporal evolution triggered by the initial perturbations in the relevant spacetimes, characterized by the complex frequencies~\cite{agr-qnm-review-01,agr-qnm-review-02,agr-qnm-review-03,agr-qnm-review-04,agr-qnm-review-06}.
As the exact solution of quasinormal modes is rather difficult to obtain analytically, in practice, one often resorts to techniques involving a certain degree of approximation.
Notably, the quasinormal modes can be reasonably evaluated through the use of the inverse P\"oschl-Teller potential~\cite{agr-qnm-Poschl-Teller-01,agr-qnm-Poschl-Teller-02}.
By adjusting the parameters of the potential to approximate that of the original black hole metric, the quasinormal frequencies are associated with the normal modes of the corresponding bound-state problem.
Similarly, the WKB method~\cite{agr-qnm-WKB-01,agr-qnm-WKB-03,agr-qnm-WKB-05} can be employed to calculate the quasinormal frequencies by using merely the derivatives of the effective potential at its maximum.
To be specific, the latter information allows to replace the potential in the region between the turning points by an approximate polynomial form.
Subsequently, the asymptotic wavefunction becomes accessible and is utilized to match others with appropriate boundary conditions in the overlap region, where the WKB approximation is valid.
In this regard, the following question arises: is it always possible to replace the effective potential of a black hole metric with an approximate form, in the sense that the underlying physics remains intact?

Intuitively, one might always argue that once a reasonably accurate approximate form is adopted for the effective potential, the resulting physics is not expected to be drastically different.
If this is not the case, any experimental measurement will be rather sensitive to the fine-tuning of the system configuration and, subsequently, the determinism of the theory is undermined.
Indeed, both examples regarding the P\"oschl-Teller and WKB methods seem to support the above heuristic arguments. 
However, subsequent studies revealed a subtlety. 
In Ref.~\cite{agr-qnm-21}, Nollert discovered a seeming contradiction by further subjecting the above considerations to a more stringent test.
By replacing the entire Regge-Wheeler potential with a series of step functions, the resulting temporal evolutions of initial perturbations as well as the quasinormal mode were examined.
The numerical calculations indicate that the time-domain waveforms are similar to those for the original smooth potential, a result in favor of the above statement.
Much to one's surprise, the resultant quasinormal modes, on the other hand, were found to be drastically different from those of the Regge-Wheeler potential.
More recently, the initial efforts of Nollert were carried on by Daghigh {\it et al.}~\cite{agr-qnm-22}.
In their study, the authors were motivated to explore whether the change in the QNM spectrum might be caused by the jump discontinuities and/or the piecewise constant nature of step functions. 
Consequently, they refined the approach by using a continuous piecewise linear potential to approximate the Regge-Wheeler potential.
It was observed, for both the staircase and linear piecewise functions, that the black hole ringdown waveform can be approximated to the desired precision by moderately increasing the number of segments.
On the other hand, the sizable difference in corresponding quasinormal frequencies persists.
These results seem to indicate that the previous findings on the asymptotic properties of the quasinormal mode spectrum are valid on general grounds.
As pointed out in Ref.~\cite{agr-qnm-21}, it leads to a rather serious question regarding the significance of black hole quasinormal modes.
To be specific, it is important to understand whether physical content carried by the quasinormal modes is distorted when the original Regge-Wheeler potential is replaced by a piecewise approximate form or, 
in other words, how to capture the essential physics of the system when such an approximation is performed.

The significance of the asymptotic behavior of quasinormal modes has been explored extensively in the literature.
Usually, for a broad class of black holes, the asymptotic quasinormal mode spectrum lines up parallel to the imaginary axis.
To be specific, the spacing of the imaginary part of the quasinormal frequency~\cite{agr-qnm-02,agr-qnm-continued-fraction-02,agr-qnm-05} is found to be $\frac{2\pi i}{\beta}$, with $\beta$ being the inverse Hawking temperature.
It is in accordance with the thermal Green's function~\cite{book-thermo-field-theory-Bellac,agr-qnm-05}. 
Its real part is speculated to be related to the quantization of black hole surface area~\cite{agr-qnm-02,agr-qnm-23}. 
The above results have been confirmed by many well-known analytic and semianalytic methods.
As discovered in Refs~\cite{agr-qnm-21,agr-qnm-22}, when the Regge-Wheeler potential is approximated by a piecewise function, the resultant quasinormal modes are found to be drastically different.
In particular, the quasinormal mode spectrum of the piecewise potential stretches out along the real axis, rather than the imaginary one.
In this context, between the results from the two potential forms, the match found in waveforms and the deviation observed in quasinormal modes pose an apparent contradiction.
Moreover, the physical nature behind the distinct asymptotic behavior of the spectra also deserves further investigation.

On the practical side, as a numerical scheme, it seems meaningful to replace the effective potential by an adequate approximate form.
As discussed above, it is expected that such a procedure will capture the essence of the physical system while substantially facilitating the numerical calculations.
However, the observed discrepancies in quasinormal modes may lead to a deeper issue associated with the numerical approach based on the finite difference method.
Usually, the algorithm works on a fixed grid.
As a result, it only utilizes potential values at a finite number of coordinate points, and thus it may not know about any sudden jump where the potential has not been sampled in practice.
To be specific, one seems to face the following ``paradox."
On the one hand, one believes that the resultant quasinormal modes will always converge to those of the Regge-Wheeler potential, once a sufficiently high resolution is adopted.
Now, let us assume that someone performs a numerical study of the temporal evolution of quasinormal oscillations for the Regge-Wheeler potential at desirably high but finite precision.
The resulting quasinormal frequencies, which may be extracted by the Prony method, are therefore expected to largely line up along the imaginary axis, close to those of the calculated black hole metric.
However, we note that during the course of the above calculations, the effective potential has only been sampled at a well-defined set of spatial grid points.
In particular, those grids furnish the only interface through which the information about the black hole metric is passed to the numerical code.
In this regard, one may devise a second effective potential consisting of a staircase piecewise function, where the values of the potential are precisely those of the Regge-Wheeler one at the grids in question.
Consequently, one proceeds by again carrying out the temporal evolution using the specifically tailored staircase ``approximate'' potential while employing an identical numerical scheme.
The latter guarantees that the same grid sites are sampled and, subsequently, the resultant quasinormal modes will be precisely the same. 
However, by employing similar arguments, the calculations using the piecewise potential are also expected to produce the quasinormal modes which line along the real axis instead.
Therefore, the drastic difference in the asymptotic behavior of quasinormal modes indicates a dilemma that might be encountered in practice.

Motivated by the above concerns, the present paper involves an attempt to shed some light on the origin of the asymptotic properties of quasinormal modes in a piecewise approximate potential.
In particular, we explore the following aspects which are still ambiguous to us at the moment.

Nollert has speculated~\cite{agr-qnm-21} that the sensitivity in the resultant quasinormal modes might be a consequence of some specific properties of the step potential.
We argue that this is indeed the case.
To be specific, we show that either a single cut or a minor step inserted into the effective potential will significantly affect the resultant quasinormal mode spectrum.
Also, to steer clear of any unnecessary complication in the vicinity of the black hole horizon, we devise our effective potential by using a somewhat different strategy from those in Refs.~\cite{agr-qnm-21, agr-qnm-22}.
Instead of replacing the entire effective potential by a piecewise function, the modifications are only placed in the region further away from the maximum of the potential.
Physically, it is also reasonable to consider such an {\it ansatz} since a discontinuity in the matter distribution might be present due to the galactic disc, space dust, or the surface of a compact object.
By using the Prony method~\cite{agr-qnm-16}, the resultant quasinormal frequencies can be extracted numerically from the temporal waveform.
We show that the asymptotic behavior of the resultant quasinormal frequencies is qualitatively consistent with those obtained by using the piecewise function approximation.
More importantly, the effective potential devised in our approach possesses the advantage of maintaining the fundamental mode. 
Besides, using the analytic approach, we demonstrate that the observed asymptotic behavior of the quasinormal mode spectrum is in agreement with the numerical results.
Also, we furnish a possible explanation of how different quasinormal spectra in the frequency domain give rise to similar waveforms in the time domain.

The derivation given in this paper is discussed and compared against other standard methods in order to understand the origin of the difference.
Last but not least, the asymptotic behavior of quasinormal modes is of physical relevance, besides being an approximation to that of a realistic black hole metric.
In particular, we argue that a piecewise potential is indeed related to several physically relevant scenarios.
As a result, the drastic difference calculated theoretically may potentially lead to implications of astrophysical relevance, which might be observed experimentally.
From the opposite perspective, precise measurements of wave propagation might offer valuable insight into the nature of the black hole spacetimes.

The plan for the rest of the paper is as follows.
In the next section, we first present the main features of the resultant quasinormal mode spectrum, by visually illustrating its distribution in the complex plane.
We explain how the modified effective potential is appropriately devised to achieve our goal and demonstrate the numerical results.
Furthermore, analytic derivations of high-overtone quasinormal modes are given in Sec.~III.
In Sec.~III~A, we study the asymptotical quasinormal frequencies in a modified P\"oschl-Teller potential.
Owing to the method proposed by Ferrari and Mashhoon~\cite{agr-qnm-Poschl-Teller-01,agr-qnm-Poschl-Teller-02}, the P\"oschl-Teller potential is both analytically tractable and captures the essence of the black hole spacetime.
In addition, another motivation involves a somewhat similar scenario explored by other authors~\cite{agr-qnm-Poschl-Teller-03,agr-qnm-Poschl-Teller-04} regarding high-overtone modes, and rather distinct results have been obtained.
Subsection III~B is devoted to the study of asymptotic quasinormal frequencies when a discontinuity is planted into the effective potential in a more general context.
In Sec.~III~C, we further generalize the results to the case when the discontinuity is present in a higher-order derivative.
In Sec.~IV, the relation with other conventional methods as well as existing results is compared and discussed. 
We discuss possible astrophysical implications of the present findings in Sec.~V. 
Further discussions, as well as the concluding remarks, are given in the last section.

\section{II. Main result and numerical calculations}

In this section, we start by schematically presenting our conclusion on the spectrum of quasinormal frequencies due to the piecewise function approximation of the potential.
The observed drastic impact on the quasinormal modes is attributed to the discontinuity brought over by the piecewise function.
In particular, we will concentrate on the effect of a single ``step'' and then demonstrate that the resultant characteristics are valid in a more general ground.
In other words, the observed feature does not depend on the specific choice regarding the specific type of discontinuity and the shape of the remaining part of the potential.
To give an instinctive notion, in the present section, we demonstrate the results obtained using the numerical approach.
More rigorous arguments and analytic derivations will be presented in Sec.~III.

\begin{figure}
\begin{tabular}{c}
\vspace{0pt}
\begin{minipage}{225pt}
\centerline{\includegraphics[width=275pt]{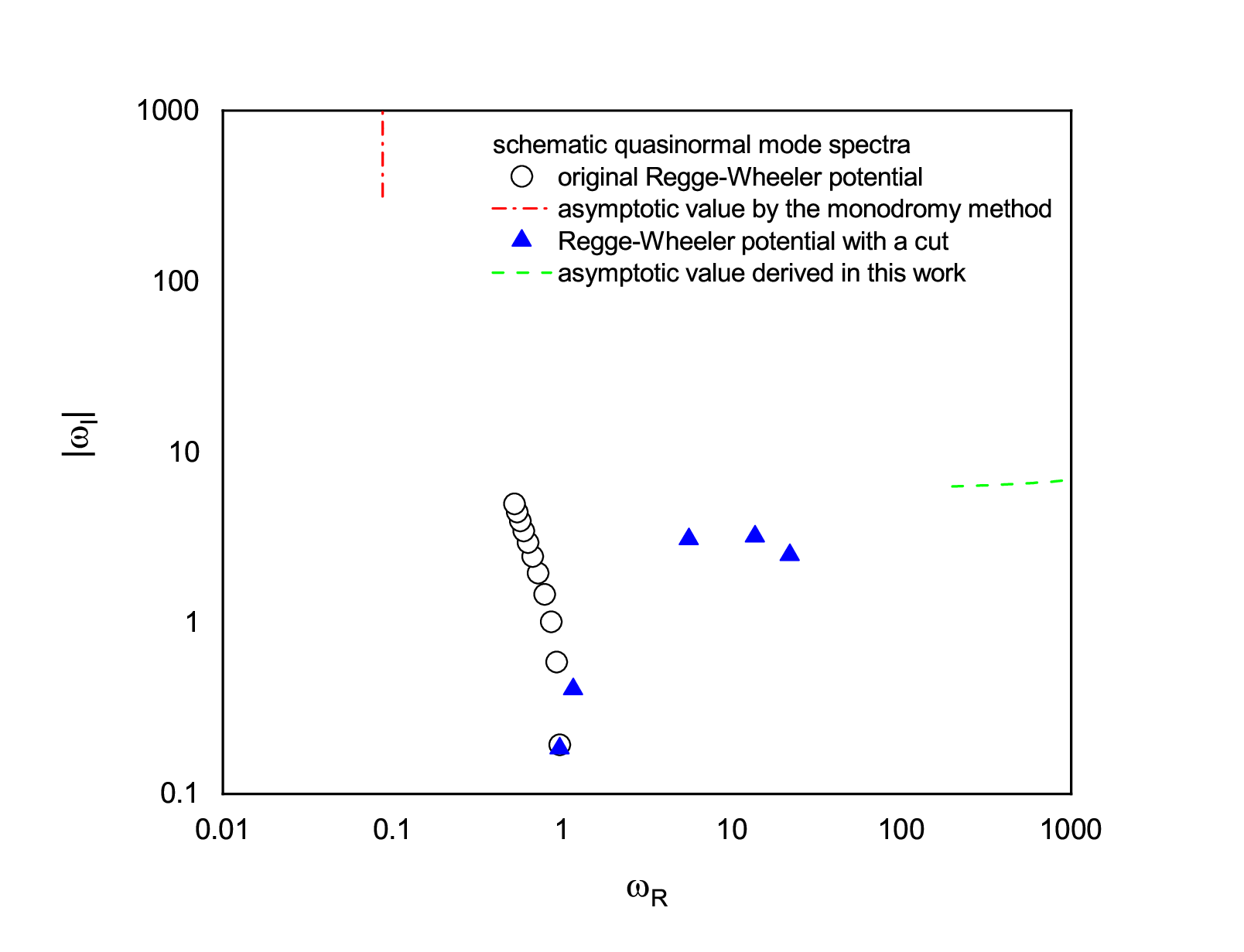}}
\end{minipage}
\end{tabular}
\caption{(Color online)
Schematic illustration of the quasinormal mode spectrum for scalar perturbations with $\ell =2$ in the complex frequency plane.
The empty black circles are the results for the Regge-Wheeler potential Eq.~\eqref{V_master} obtained using the continued fraction method.
The red dash-dotted line parallel to the imaginary axis indicates the corresponding asymptotic behavior with a large imaginary part.
The filled blue triangles represent the most dominant quasinormal frequencies when the tail of the Regge-Wheeler potential is cut off at $r=10$, as explicitly given in Eq.~\eqref{V_cut}.
The corresponding asymptotic behavior of the quasinormal modes is given by the green dashed curve.
}
\label{schematic_qnm}
\end{figure}

In practice, the study of black hole perturbation theory involves the solution of the radial part of the master equation~\cite{agr-qnm-review-03,agr-qnm-review-06},
\begin{eqnarray}
\frac{\partial^2}{\partial t^2}\Psi(t, x)+\left(-\frac{\partial^2}{\partial x^2}+V\right)\Psi(t, x)=0 ,
\label{master_eq_ns}
\end{eqnarray}
where the effective potential $V$ is determined by the given spacetime metric, spin ${\bar{s}}$, and angular momentum $\ell$ of the perturbation.
For instance, the Regge-Wheeler potential for the Schwarzschild black hole metric reads
\bqn
V_\mathrm{RW}=f\left[\frac{\ell(\ell+1)}{r^2}+(1-{\bar{s}}^2)\frac{r_h}{r^3}\right],
\lb{V_master}
\eqn
where 
\bqn
f=1-r_h/r ,
\lb{f_master}
\eqn
where the horizon $r_h=2M$ and $M$ is the mass of the black hole.
For simplicity, in the present study, we consider the scalar ($s=0$) and vector-type gravitational ($s=2$) perturbations with $r_h=1$ and the multiple index $\ell=2$.

As discussed above, our main goal is to investigate the drastic modification of the quasinormal mode spectrum due to the discontinuity in the piecewise function. 
In Refs.~\cite{agr-qnm-21,agr-qnm-22} remarkable distinctions were found in the quasinormal frequencies for both the fundamental modes and asymptotic behavior.
To proceed, our strategy is to separate the essential cause of the above two features.
To start with, let us consider a scenario where only a single ``cut'' or ``step'' is introduced to the Regge-Wheeler potential, and is located further away from both the initial perturbations and an observer.
As dictated by causality, it is expected that the waveform measured by the observer will be precisely the same as those for the original Regge-Wheeler potential, until the signal reaches the discontinuity and partly bounces back to the observer.
As a result, even if the overall spectrum is dramatically affected for some reason, the fundamental mode should remain largely unchanged.
In such a way, one manages to isolate the factor which is crucial to the asymptotic behavior of the quasinormal modes from those related to the properties of the spacetime in the vicinity of the horizon.
Regarding the fundamental mode (as will also be further discussed) the deviation from that of the Regge-Wheeler potential can be understood as the potential and its derivatives will be significantly distorted if any discontinuity is located in the vicinity of its maximum.

In Fig.~\ref{schematic_qnm} we present the resultant schematic distribution of the quasinormal modes in the complex frequency plane.
The spectrum of the quasinormal modes of the scalar perturbations is shown for both the original Regge-Wheeler potential and the one with a discontinuity.
For the latter, as discussed above, we consider the case where the tail of the potential is cut off at $r=10$.
The low-lying modes for the original Regge-Wheeler potential are shown as empty black circles, obtained using the continued fraction method~\cite{agr-qnm-continued-fraction-01}.
The red dash-dotted line parallel to the imaginary axis indicates the asymptotic behavior with large imaginary part~\cite{agr-qnm-02,agr-qnm-continued-fraction-02,agr-qnm-continued-fraction-02,agr-qnm-05}.
For the modified Regge-Wheeler potential with a cut, although the initial temporal evolution remains unchanged, the low-lying modes (as shown by the filled blue triangles) are significantly different.
The values of these modes obtained by evaluating the temporal evolution of the initial oscillation using the finite difference method and then extracting the dominant complex frequencies by employing the Prony method.
Indeed, apart from the fact that the fundamental mode mostly coincides with that from the original potential, the distribution of the quasinormal modes is along the real axis, reminiscent of those obtained in Refs.~\cite{agr-qnm-21,agr-qnm-22}.
Moreover, the adopted configuration facilitates the study of quasinormal frequencies with large real and imaginary parts.
The asymptotic behavior of the quasinormal mode spectrum is indicated by the green dashed curve.
The derivation of the latter will be given in the next section.

\begin{figure}
\begin{tabular}{cc}
\vspace{0pt}
\begin{minipage}{225pt}
\centerline{\includegraphics[width=200pt]{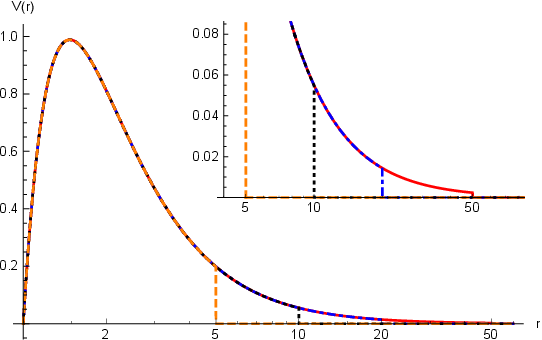}}
\end{minipage}
&
\begin{minipage}{225pt}
\centerline{\includegraphics[width=200pt]{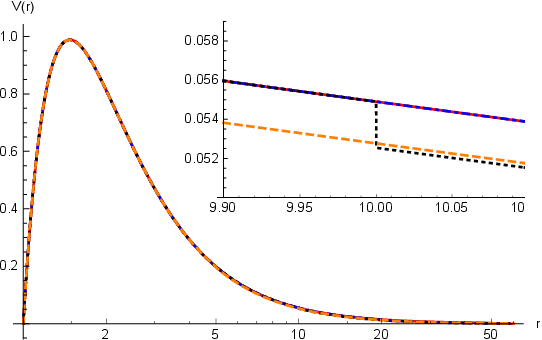}}
\end{minipage}
\end{tabular}
\renewcommand{\figurename}{Fig.}
\caption{(Color online)
Modified potential based on the Regge-Wheeler one investigated in the present study.
The left and right plots show the two groups of potentials given in Eqs.~\eqref{V_cut} and~\eqref{V_step}.
The orange dashed, black dotted, blue dot-dashed, and red solid curves correspond to a cut or step introduced at $r=5, 10, 20$, and $50$ respectively.}
\label{toy_potential}
\end{figure}

One might doubt whether the observed feature of the low-lying quasinormal modes for the modified Regge-Wheeler potential presented in Fig.~\ref{schematic_qnm} merely corresponds to a particular case.
The remainder of this section will be devoted to this issue.
To be specific, we argue that the number, size, location, and even the order of the discontinuity are not determinant factors.
First of all, our calculations indicate that a single continuity suffices to reproduce the spectrum which lines up along the real axis; we will continue to concentrate on this simple choice throughout this paper.
In order to show that the feature observed above is indeed robust, we will explicitly deal with other possible modifications to the Regge-Wheeler potential.
In particular, we consider two groups of parametrizations.
The first group, similar to the one demonstrated in Fig.~\ref{schematic_qnm}, is devised by cutting off the tail of the Regge-Wheeler potential from a given radius $r_{\mathrm{cut}}$, namely,
\bqn
V_{\mathrm{cut}}=
\left\{\begin{array}{cc}
V_\mathrm{RW}(r)    &  r\le r_{\mathrm{cut}}  \cr\\
0&  r> r_{\mathrm{cut}}  
\end{array}\right. , 
\lb{V_cut}
\eqn
The resultant modified potential is investigated by taking various values of $r_{\mathrm{cut}}$, as shown in the left plot of Fig.~\ref{toy_potential}.

\begin{figure}
\begin{tabular}{cc}
\vspace{0pt}
\begin{minipage}{225pt}
\centerline{\includegraphics[width=200pt]{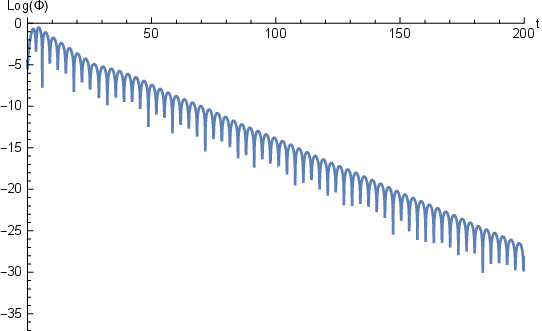}}
\end{minipage}
&
\begin{minipage}{225pt}
\centerline{\includegraphics[width=200pt]{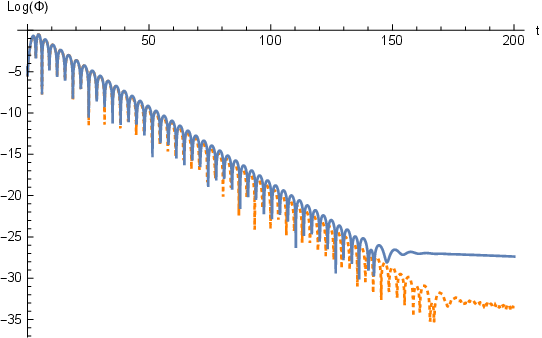}}
\end{minipage}
\end{tabular}
\renewcommand{\figurename}{Fig.}
\caption{(Color online)
Resultant temporal evolutions measured by an observer located at $r_*=0$ for some initial Gaussian distribution centered at $r_*=3$ with a width $\sigma=1$.
The blue solid curves shown in the left and right plots correspond to the cases where the ``cut'' and ``step'' defined in Eqs.~\eqref{V_cut} and~\eqref{V_step} are located at $r_*\sim r=10$.
While the initial oscillations for $t \lesssim 15$ are identical for the two cases, a late-time tail for $t \gtrsim 150$ is observed for the potential with the ``step".
As a comparison, the results for the original Regge-Wheeler potential are also represented by the orange dotted curves in the right plot.
}
\label{toy_evlution}
\end{figure}

\begin{figure}
\begin{tabular}{cc}
\vspace{0pt}
\begin{minipage}{225pt}
\centerline{\includegraphics[width=275pt]{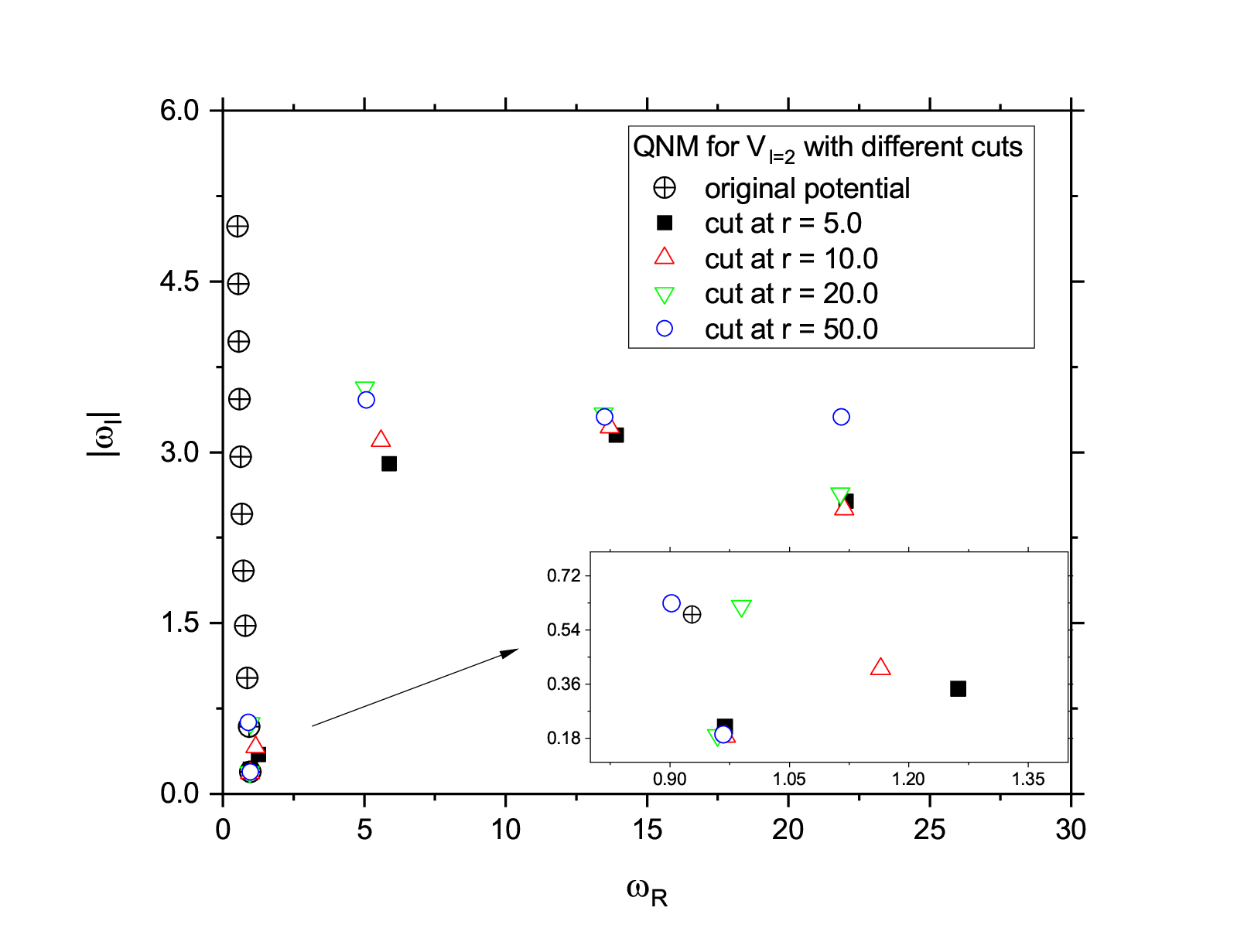}}
\end{minipage}
&
\begin{minipage}{225pt}
\centerline{\includegraphics[width=275pt]{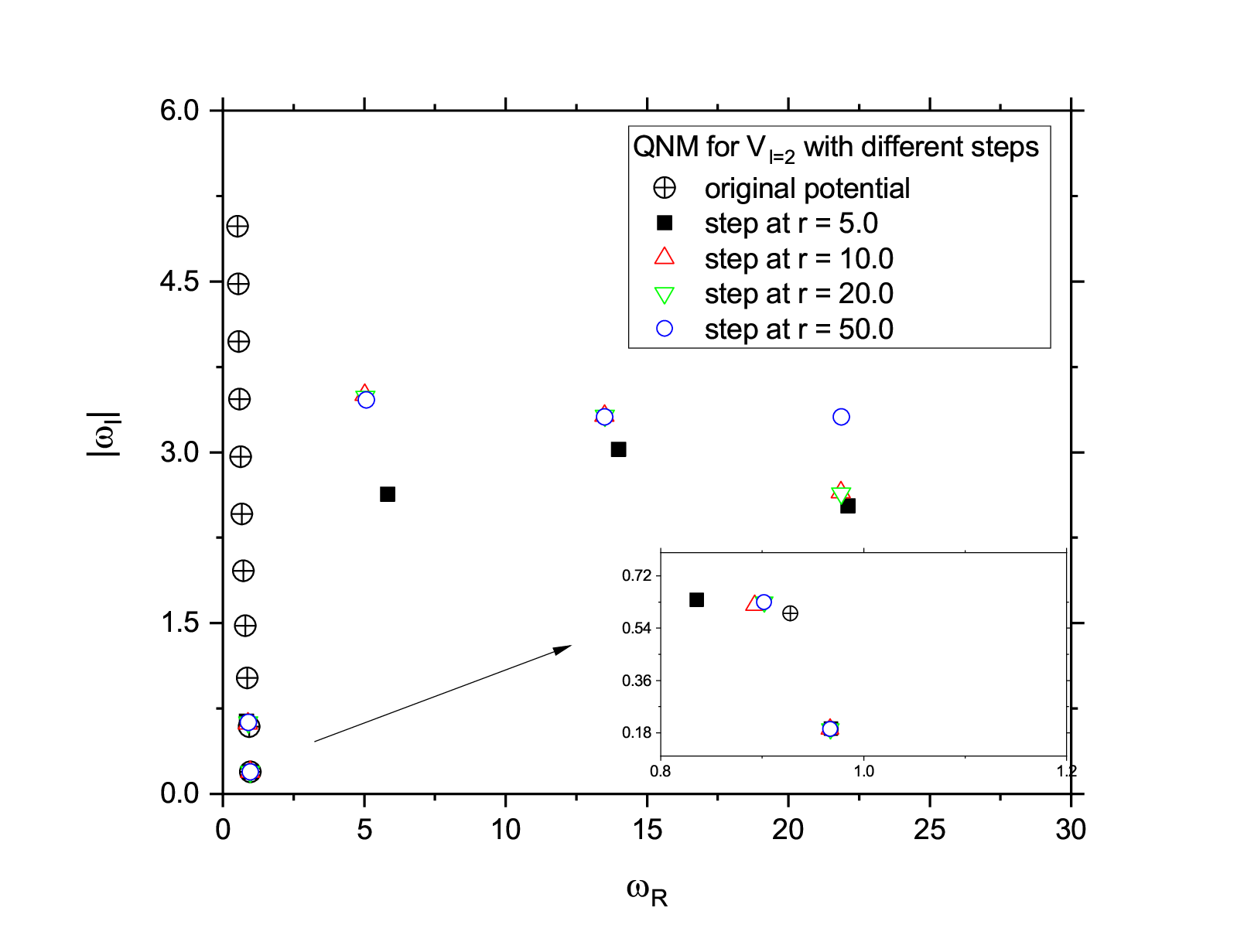}}
\end{minipage}
\end{tabular}
\renewcommand{\figurename}{Fig.}
\caption{(Color online)
Resultant low-lying quasinormal modes obtained from the potentials given in Fig.~\ref{toy_potential}.
The left and right plots show the two groups of potentials given in Eqs.~\eqref{V_cut} and~\eqref{V_step}.
}
\label{toy_qnm}
\end{figure}

The temporal oscillations presented in Fig.~\ref{toy_evlution} are obtained by solving Eq.~\eqref{master_eq_ns} using the finite difference method implemented in the tortoise coordinates
\bqn
r_*=r+r_h \ln\left(\frac{r}{r_h}-1\right) . 
\lb{x_tortoise}
\eqn
Subsequently, the quasinormal modes are extracted using the Prony method.
The first five primary modes are shown in the left plot of Fig.~\ref{toy_qnm} and listed in Tab.~\ref{tb_qnm_comp}. 
In Tab.~\ref{tb_qnm_comp}, we present the results for both scalar and vector-type gravitational perturbations.
The complex frequencies extracted using the Prony method are ordered by their respective importance in constituting the waveform.
For both cases, one observes that these low-lying quasinormal modes are distributed along the real axis.
It is noted that this feature is found largely independent of the position of the cut.
Also, as expected, when the location of the cut moves further away from the horizon, the fundamental mode becomes closer to that of the original Regge-Wheeler potential.

In our calculations, the temporal evolution is carried out by using the first-order finite difference method by taking the grid size $\Delta r_*=2\Delta t=0.1$, and the Prony method is carried out for the interval $(10,200)$.
The precision of the code is tuned until the extracted frequencies of the first three overtone modes appropriately match those of the Regge-Wheeler potential, which are also shown in Tab.~\ref{tb_qnm_comp}.
In the case of the modified Regge-Wheeler potential, however, when using the same settings, the resultant modes with larger real parts are not as precise.
This is somewhat expected as the accuracy of the method decreases when the period of the oscillations decreases.
Therefore, for modes of higher overtone numbers, the performance of the code suffers more significantly when the quasinormal modes line up along the real axis instead of the imaginary one.  
Moreover, when compared with the case of the original Regge-Wheeler potential, we notice that the subtracted frequencies are more sensitive to the range where the fit of the Prony method is carried out.
Nonetheless, the observed tendency of the quasinormal frequencies always persists, and the results here will be fortified further by the derivation given in the next section.

The ``cut'' investigated above might introduce a substantial discontinuity in the potential, especially when $r_{\mathrm{cut}}$ is not so large. 
To show that the strength of the discontinuity is actually an irrelevant factor, we consider below the second group of potentials.
These potentials feature a minor ``cut'' and the tail is maintained.
An explicit definition is given in Eq.~\eqref{V_step}. 
As shown in the right plot of Fig.~\ref{toy_potential}, visually, the modification to the original potential is rather insignificant when compared with the left plot.

\bqn
V_{\mathrm{step}}=
\left\{\begin{array}{cc}
V_\mathrm{RW}(r)     &  r\le r_{\mathrm{step}}  \cr\\
V_\mathrm{RW}(r)-\epsilon \exp[-r/L]     &  r> r_{\mathrm{step}}  
\end{array}\right. ,
\lb{V_step}
\eqn
where in our calculations we take $\epsilon=V_\mathrm{RW}(L)$ and $L=50$.

By following a similar procedure, the temporal evolution is shown in the right plot of Fig.~\ref{toy_evlution}.
As one compares the blue solid curve against that in the left plot, it is observed that the initial oscillations measured by the observer for $t \lesssim 15$ are identical for the two cases.
This is understood since any distinct feature demonstrated by the two potentials must also comply with the requirement of causality.
Moreover, a late-time tail starting at $t \sim 150$ is observed for the potential with the ``step", as its associated effective potential contains an appropriate tail~\cite{agr-qnm-tail-06}.

From the temporal evolution, one again obtains the corresponding low-lying quasinormal frequencies as presented in Tab.~\ref{tb_qnm_comp} and the right plot of Fig.~\ref{toy_qnm}.
By comparing the right plot of Fig.~\ref{toy_qnm} to the left one, it is straightforward to observe that the quasinormal mode spectrum associated with a ``step'' bears a strong resemblance to that with a ``cut".
In other words, we have shown numerically that the strength of the discontinuity plays a minor role in determining the quasinormal modes.

As a reference, we also show the results obtained using the original Regge-Wheeler potential.
From the right plot of Fig.~\ref{toy_evlution}, one can see that the time evolutions of the two cases are largely similar.
Again, the difference is not significant, and it is more pronounced at late times, especially when the power-law decay takes place.
However, as one takes a closer look at the resultant modes (as presented in Tab.~\ref{tb_qnm_comp}), the quasinormal modes show a more substantial difference.
For the Regge-Wheeler potential, using the Prony method, the first three dominant quasinormal modes are in reasonably good agreement with those from the standard continued fraction method.
However, the weights of the two other modes are rather insignificant compared to the others.
Therefore, it seems that their deviations from the results obtained using the continued fraction method can be attributed entirely to the numerical uncertainties. 
On the other hand, the amplitudes of the modes extracted from the modified Regge-Wheeler potentials decrease gradually with increasing overtone number.
Subsequently, by and large, all of these modes will contribute to the waveform at the initial stage.
Moreover, as the imaginary parts of the modes in question are of the same order of magnitude, they are expected to persist in the waveform over an extended period.
This is quite different from the case of the original Regge-Wheeler potential, where the fundamental mode stands out among others and dominates the time evolution.
To a certain degree, the above viewpoint might alleviate the difficulties in understanding why drastically different quasinormal mode spectra in frequency space are associated with almost identical waveforms in the time domain.
To be specific, one should not compare one dominant quasinormal mode against another, but rather compare its waveform with that formed by a summation of similarly weighted components.
Rather than a continuous spectrum, it consists of a superposition of temporal oscillations with discrete frequencies.
But if these quasinormal modes furnish a complete set, the summation in question is nothing but an expansion uniquely determined by the appropriate boundary conditions. 

Nonetheless, by evaluating only a few modes using the Prony method, it is still not convincing that the entire spectrum would universally line up alone the real axis.
Also, we have not addressed the impact of a high-order discontinuity in the potential.
These issues will be covered in the following sections, where we proceed to study the properties of the quasinormal mode spectrum with a large frequency.

\begin{table}[htb]
\begin{center}
\scalebox{1.00}
{\begin{tabular}{|c|c|c|c|c|c|c|c|}
\hline
& \multicolumn{6}{c|}{Prony method} & {continued fraction} \\ \cline{2-8}  
\multirow{2}{*}{n}& \multicolumn{2}{c|}{~$V_{\mathrm{cut}} $~} & \multicolumn{2}{c|}{~$V_{\mathrm{step}} $~} & \multicolumn{2}{c|}{~$V_{\mathrm{RW}}$~} & \multicolumn{1}{c|}{~$V_{\mathrm{RW}}$~} \\
& \multicolumn{2}{c|}{~$(\ell=2, \bar{s}=0, r_{\mathrm{cut}}=10)$~} & \multicolumn{2}{c|}{~$(\ell=2, \bar{s}=0, r_{\mathrm{step}}=10)$~} & \multicolumn{2}{c|}{~$(\ell=2, \bar{s}=0)$~} & \multicolumn{1}{c|}{~$(\ell=2, \bar{s}=0)$~} \\ \hline
$0$& ~$ 0.9709 - 0.1853i$~ &$10^{-2}$& ~$0.9673 - 0.1936i$~ &$10^{-2}$& ~$0.9673-0.1934i$~ &$10^{-2}$& $0.967288 - 0.193518i$    \\  \hline
$1$ & ~$1.165 - 0.4104i$~ &$10^{-3}$& ~$0.8926 - 0.6192i$~  &$10^{-3}$& ~$0.9262 - 0.5912i$~ &$10^{-3}$& $0.927701 - 0.591208i$   \\  \hline
$2$& ~$5.675 - 3.102i $~ &$10^{-3}$& ~$5.017 - 3.506i $~ &$10^{-6}$& ~$0.8618 -1.099i $~ &$10^{-4}$& $0.861088 - 1.017117i$   \\  \hline
$3$ & ~$13.79 - 3.222i $~ &$10^{-4}$& ~$13.51 - 3.324i $~ &$10^{-6}$& ~$23.23-1.171i$~ &$10^{-11}$& $0.787726 - 1.476193i$  \\  \hline
$4$ & ~$21.99 - 2.599i $~ &$10^{-5}$& ~$21.86 - 2.640i $~ &$10^{-6}$& ~$31.42-1.234i$~ &$10^{-11}$& $0.722598 - 1.959843i$  \\  \hline
\multirow{2}{*}{n}& \multicolumn{2}{c|}{~$V_{\mathrm{cut}} $~} & \multicolumn{2}{c|}{~$V_{\mathrm{step}} $~} & \multicolumn{2}{c|}{~$V_{\mathrm{RW}}$~} & \multicolumn{1}{c|}{~$V_{\mathrm{RW}}$~} \\
& \multicolumn{2}{c|}{~$(\ell=2, \bar{s}=2, r_{\mathrm{cut}}=10)$~} & \multicolumn{2}{c|}{~$(\ell=2, \bar{s}=2, r_{\mathrm{step}}=10)$~} & \multicolumn{2}{c|}{~$(\ell=2, \bar{s}=2)$~} & \multicolumn{1}{c|}{~$(\ell=2, \bar{s}=2)$~} \\ \hline
$0$& ~$ 0.7590 - 0.1808i$~ &$10^{-2}$& ~$0.7476 - 0.1784i$~ &$10^{-2}$& ~$0.7474-0.1779i$~ &$10^{-2}$& $0.747343 - 0.177925i$    \\  \hline
$1$ & ~$0.9709 - 0.3289i$~ &$10^{-3}$& ~$0.6889 - 0.5962i$~  &$10^{-4}$& ~$0.6928 - 0.5465i$~ &$10^{-4}$& $0.693422 - 0.547830i$   \\  \hline
$2$& ~$5.662 - 3.222i $~ &$10^{-4}$& ~$4.719 - 3.930i $~ &$10^{-6}$& ~$0.4917 -0.9949i $~ &$10^{-4}$& $0.602107 - 0.956554i$   \\  \hline
$3$ & ~$13.92 - 3.222i $~ &$10^{-4}$& ~$13.51 - 3.383i $~ &$10^{-6}$& ~$40.35-2.443i$~ &$10^{-10}$& $0.503010 - 1.410296i$  \\  \hline
$4$ & ~$22.07 - 2.577i $~ &$10^{-5}$& ~$21.89 - 2.640i $~ &$10^{-7}$& ~$55.53-2.666i$~ &$10^{-12}$& $0.415029 - 1.893690i$  \\  \hline
\end{tabular}}
\end{center}
\caption{Calculated scalar and vector-type gravitational quasinormal frequencies.
The calculations are carried out for the effective potentials defined in Eqs.~\eqref{V_cut} and \eqref{V_step}, as well as the original Regge-Wheeler potential Eq.~\eqref{V_master}.
From left to right, the results shown in the first three columns are obtained using the Prony method extracted from the temporal evolutions as shown in Fig.~\ref{toy_evlution}.
The more accurate values of the quasinormal frequencies for the potential $V_{\mathrm{RW}}$ obtained using the continued fraction method are also given in the fourth column as a reference.
Also, the results obtained using the Prony method are ordered by their respective weights in the waveform, which are also presented next to the values of the quasinormal frequencies.
}
\label{tb_qnm_comp}
\end{table}

\section{III. Asymptotic properties of quasinormal mode spectrum in piecewise effective potential}

In this section, we explore the asymptotic properties of the quasinormal mode spectrum.
In Sec.~III~A, we first devise a specific simple problem that can be solved more straightforwardly.
We move to a more general proof in Sec.~III~B, which is largely based on the proper treatment of the discontinuity where the WKB approximation breaks down.
In Sec.~III~C, the results are further generalized to the cases where the discontinuity is only present at a higher order. 

\subsection{A. An explicit example with the modified P\"oschl-Teller potential}

In this subsection, we construct an effective potential that can be primarily treated analytically, which to some certain degree reflects a realistic problem.
Similar to the effective potential given above in Eq.~\eqref{V_cut}, we introduce a ``cut'' in the P\"oschl-Teller potential located at a radius further away from the maximum of the potential.
Since the solution for the P\"oschl-Teller potential is known analytically, it is feasible to study the asymptotic behavior of the quasinormal mode spectrum analytically.

Before proceeding further, we point out a subtlety in the derivation.
To explore the asymptotic properties of the quasinormal frequencies due to a cut located spatially distant from the horizon, one has to deal with two limits.
To be specific, the location of the cut approaches spatial infinity, $x_\mathrm{cut}\to +\infty$.
Meanwhile, the real part of the frequency also goes to infinity, $\mathrm{Re}\omega\equiv \omega_R\to +\infty$.
However, if one takes the limit for the cut first, namely, $\lim\limits_{\omega_R\to +\infty}\lim\limits_{x_\mathrm{cut} \to +\infty}$, (as will become obvious) the ``cut'' will not play any role in the final result.
Instead, during the course of the derivation, one should assume that the ``cut'' is located at a relatively large but finite radius.
One may take the limit of asymptotic spatial infinity only at the end of the calculation in order to discuss the physical relevance of the obtained results.

In what follows, we derive the quasinormal frequencies by evaluating the zeros of the Wronskian determinant
\begin{eqnarray}
W(\omega)\equiv W(g,f)=g(\omega,x)f'(\omega, x)-f(\omega,x)g'(\omega,x) ,
\label{pt_Wronskian}
\end{eqnarray}
where $'\equiv d/dx$.
Here $f$ and $g$ are the solutions of the homogenous equation in s-domain~\cite{agr-qnm-review-02},
\begin{eqnarray}
\left[-\omega^2-\frac{d^2}{dx^2}+\tilde{V}_\mathrm{PT}\right]\tilde{\Psi}=0 ,
\label{pt_homo_eq}
\end{eqnarray}
with appropriate boundary conditions, namely,
\begin{eqnarray}
\begin{array}{cc}
f(\omega, x)\sim e^{-i\omega x}    &  x\to -\infty  \cr\\
g(\omega, x)\sim e^{i\omega x}     &  x\to +\infty  
\end{array} ,
\label{pt_boundary}
\end{eqnarray}
The effective potential 
\bqn
\tilde{V}_\mathrm{PT}=
\left\{\begin{array}{cc}
{V}_\mathrm{PT}(x)     &  x\le x_{\mathrm{cut}}  \cr\\
0    &  x> x_{\mathrm{cut}}  
\end{array}\right. ,
\lb{V_mpt}
\eqn
is constructed by introducing a ``cut'' at $x_{\mathrm{cut}} $ in the P\"oschl-Teller potential,
\begin{eqnarray}
{V}_\mathrm{PT}=\frac{V_m}{\cosh ^2(\kappa x)} .
\end{eqnarray}

To proceed, we first obtain $f$ and $g$ and then evaluate the zeros of the Wronskian Eq.~\eqref{pt_Wronskian} to find the quasinormal frequencies.
Based on the well-known method~\cite{agr-qnm-Poschl-Teller-01,agr-qnm-Poschl-Teller-02}, the quasinormal modes for the original (inverse) P\"oschl-Teller potential can be directly obtained from the eigenvalues of the corresponding bound-state problem.
Now, for the modified potential Eq.~\eqref{V_mpt}, one cannot straightforwardly apply the method.
However, using an adapted procedure, the wave function $f(\omega,x)$ satisfying the first line of Eq.~\eqref{pt_boundary} regarding the potential $\tilde{V}_\mathrm{PT}$ can be obtained similarly by introducing the transformation 
\bqn
\left\{\begin{array}{c}
x\to -ix  \cr\\
\kappa \to i\kappa  
\end{array}\right. ,
\lb{trans_PT}
\eqn
while making proper combinations of the analytic forms of the solutions with well-defined parities~\cite{book-quantum-mechanics-Flugge}.
The resultant form of $f(\omega,x)$ can be written as
\bqn
f(\omega,x) = A u_e + B u_o . 
\lb{f_mPT}
\eqn
We delegate the specific forms of $u_e, u_o$ together with the straightforward but somewhat tedious derivations of the coefficients $A, B$ to Eqs.~\eqref{ueuo_solution}-\eqref{AB_ueuo} in the Appendix.
Here we only note that, as coefficients, $A$ and $B$ are not functions of the tortoise coordinate $x$.

On the other hand, the resultant form of the wave function $g(\omega, x)$ satisfying the boundary condition Eq.~\eqref{pt_boundary} can be easily obtained,
\bqn
g(\omega,x) = e^{i\omega x} . 
\lb{g_mPT}
\eqn

Subsequently, the quasinormal frequencies can be obtained by the requirement that the Wronskian Eq.~\eqref{pt_Wronskian} vanishes.
Now, it is interesting to point out that if one takes the limit $x_\mathrm{cut}\to +\infty$ in the wave functions $u_e, u_o$ at this moment and then substitutes Eq.~\eqref{f_mPT} into the Wronskian,
one finds
\bqn
f(\omega,x) \to C e^{i\omega x} + D e^{-i\omega x} \ \ \ \mathrm{for}\ x\to+\infty, 
\lb{f_mPT_CD}
\eqn
with $C, D$ given by Eq.~\eqref{CD_ueuo}.
The condition for the Wronskian to vanish therefore occurs at the poles of $C/D$, which corresponds to the case of a pure outgoing wave at $x\to +\infty$.
By explicitly making use of Eq.~\eqref{CD_ueuo}, it corresponds to the condition for any factor of the product $\Gamma(b)\Gamma\left(b+\frac12\right)\Gamma\left(1-a\right)\Gamma\left(\frac12-a\right)$ to diverge.
Subsequently, the quasinormal frequencies are obtained by substituting Eqs.~\eqref{trans_PT} and~\eqref{trans_PT_omega}, which are
\bqn
\omega^{\mathrm{PT}}_R &=& \sqrt{V_m-\frac{\kappa^2}{4}},\nonumber \\
\omega^{\mathrm{PT}}_I &=& -\left(n+\frac12\right)\kappa ,
\lb{qnm_PT}
\eqn
where $n$ is a non-negative integer.
Since $C/D$ is nothing but the ratio of the amplitude of the reflected wave to the incident one, the above procedure is precisely what the method dictates~\cite{agr-qnm-Poschl-Teller-01,agr-qnm-Poschl-Teller-02}.
It is noted that the quasinormal mode spectrum climbs up along the imaginary axis.

As mentioned above, what will radically change the result is that a ``cut'' is implemented at a finite location.
In particular, the Wronskian is no longer evaluated at asymptotic spatial infinity, but at $x_\mathrm{cut}$. 
Such a modification leads to a small additional contribution due to the deviation from the poles of the first line of Eq.~\eqref{CD_ueuo}.
In other words, for the original quasinormal frequency, the wave function $f(\omega, x)$ at $x=x_\mathrm{cut}$ is not purely outgoing but contains a small fraction of the ingoing wave, and therefore the requirement that the Wronskian must vanish indicates a correction to the frequency.
For an arbitrary frequency, the condition for the Wronskian to vanish at the ``cut'' implies that the small deviations from the asymptotical outgoing wave precisely cancel out those related to the ingoing one. 
We argue that this novel condition eventually leads to a dramatic change in the quasinormal mode spectrum.
Although the above arguments seem to be rather heuristic and only apply to a particular example, they will become more transparent in the following subsection where many key features discussed here are found to be general.

Instead of evaluating the Wronskian entirely, one only needs to calculate the difference compared to its value for the P\"oschl-Teller potential.
As we are more interested in the asymptotic properties of the quasinormal modes, the above difference can be estimated by expanding the wave function at $x\to +\infty$ to the second order. 
By making use of the coefficients derived in Eqs.\eqref{CD_tilde_ueuo}-\eqref{CD_tilde_delta_ueuo}, the resultant deviation reads
\bqn
W(\omega)= \Delta W(\omega) = \Delta W_C(\omega) + \Delta W_D(\omega). 
\lb{Wronskian_PT_dev}
\eqn
where 
\bqn
\Delta W_C(\omega) \doteq W(g, \Delta C e^{i\omega x}) = -2\kappa (\Delta\tilde{C}_1+\Delta\tilde{C}_2) e^{-2\kappa x_\mathrm{cut}}e^{2i\omega x_\mathrm{cut}}
\lb{Wronskian_PT_dev_C}
\eqn
and 
\bqn
\Delta W_D(\omega) \doteq W(g, D e^{-i\omega x}) = -2i\omega D
\lb{Wronskian_PT_dev_D}
\eqn

Equating Eq.~\eqref{Wronskian_PT_dev} to zero gives an algebraic equation for the quasinormal frequencies $\omega$.
At the limit $\omega \to \infty$, assisted by the asymptotic relation Eq.~\eqref{DeltaCD_ratio} derived in the Appendix, the equation takes the simple form
\bqn
\frac{\omega^2}{V_m}=e^{-2\kappa x_\mathrm{cut}} e^{2i\omega x_\mathrm{cut}} .
\lb{qnm_Eq_mPT}
\eqn
We note that since the equation involves the term $e^{2i\omega x_\mathrm{cut}}$, the real part of $\omega$ can be shifted since $e^{2i\omega x_\mathrm{cut}}=e^{2i\omega x_\mathrm{cut}+i2\pi j}$, with $j$ being an (in particular, arbitrarily large) integer.
Subsequently, the asymptotic form of the quasinormal frequencies is found to be
\bqn
\omega_R &=& \frac{\pi j}{x_\mathrm{cut}}+O\left(1\right) ,\nonumber \\
\omega_I &=& -\frac{\ln(\pi j)}{x_\mathrm{cut}}-\kappa +\frac{\ln x_\mathrm{cut}}{x_\mathrm{cut}}+\frac{\ln V_m}{2x_\mathrm{cut}}+O\left(1\right) ,
\lb{qnm_mPT}
\eqn
where $j$ is a (large) integer.
In accordance with the numerical results, the quasinormal frequencies now line up along the real axis.
As discussed above, the poles of the Gamma function give rise to quasinormal frequencies located close to the imaginary axis.
Therefore, when a ``cut'' is introduced, the relevant frequencies steer clear of these poles.

Also, if the location of the ``cut'' is further away from the horizon, as $x_\mathrm{cut}$ increases, the real part of the frequency for a given overtone number becomes smaller.
However, the asymptotic behavior does not change.
This confirms what we have claimed above.
To be specific, if the limit $x_\mathrm{cut}\to +\infty$ is taken at the end of the calculation, the conclusion remains unchanged.
The resultant quasinormal modes given in Eq.~\eqref{qnm_mPT} are indeed drastically different from Eq.~\eqref{qnm_PT}.

\subsection{B. General results for a potential with discontinuity}

In this subsection, we provide a more general derivation of the above results.
We consider a potential $V(x)$ defined in tortoise coordinates that vanishes asymptotically as $x \to \infty$.
Again, for simplicity, a single ``cut'' is introduced to the potential at the position $x_\mathrm{cut} (> 1)$ similar to the case of Eq.~\eqref{V_mpt}.

As the potential is cut off beyond $x=x_\mathrm{cut}$, it is obvious that the solution Eq.~\eqref{g_mPT}, which satisfies the second line of Eq.~\eqref{pt_boundary}, remains valid on the r.h.s. of the discontinuity, 
To deal with the general form of potential on the l.h.s. of the ``cut", we assume that the WKB approximation~\cite{book-quantum-mechanics-Sakurai} is valid.
In other words, the wave function can be obtained by the formal integration
\bqn
f(\omega, x)=C e^{iS(x_0,x)}+D e^{-iS(x_0,x)} ,
\lb{wf_WKB}
\eqn
where, at the lowest-order approximation (which suffices for the present case), we have
\bqn
S(x_0,x)=\int_{x_0}^{x} k(x')dx' ,
\lb{S_WKB}
\eqn
where $k(x)=\sqrt{\omega^2-V(x)}$, and $x_0$ is taken somewhat arbitrarily in the region where the WKB formula is valid.

As already learned from a particular example, the coefficients $C$ and $D$ are not arbitrary, as they are determined in order to guarantee the outgoing boundary condition Eq.~\eqref{pt_boundary}.
The evaluation of these coefficients is not trivial as they are related to the specific form of the potential.
Fortunately, as shown below, under moderate assumptions their particular forms are not crucial to the present discussion.
What is relevant is that $C$ and $D$ are well defined in terms of the potential as well as the boundary condition of the wave function at $x\to -\infty$.

The quasinormal mode spectrum can be obtained by evaluating the Wronskian.
Again, if one first takes the limit $x_\mathrm{cut}\to +\infty$, the potential vanishes asymptotically, $S(x)\to i\omega x$.
Subsequently, the quasinormal modes would be dictated by the poles of the ratio $C/D$, which corresponds to the conventional scenario.

On the contrary, for any finite $x_\mathrm{cut}$, it is crucial that $D\ne 0$ at $x_\mathrm{cut}$, precisely because of the presence of the discontinuity.
As $x=x_\mathrm{cut}$ is the point where the WKB approximation breaks down, one has to resort to the Wronskian to connect the wave function on both sides of the discontinuity.
To be specific,
\bqn
W(\omega)=e^{i\omega x_\mathrm{cut}}(ik)(Ce^{iS_\mathrm{cut}}-De^{-iS_\mathrm{cut}})-(Ce^{iS_\mathrm{cut}}+De^{-iS_\mathrm{cut}})(i\omega)e^{i\omega x_\mathrm{cut}}=0 ,
\lb{WKB_Wronskian_1st}
\eqn
where $S_\mathrm{cut} = S(x_0, x_\mathrm{cut})$.
For large $\omega$, it is straightforward to find
\bqn
\omega_R &=& \frac{\pi j}{x_\mathrm{cut}} + O(1) ,\nonumber \\
\omega_I &=& -\frac{\ln(\pi j)}{x_\mathrm{cut}}+\frac{\ln x_\mathrm{cut}}{x_\mathrm{cut}} - \frac{\ln(D/C)}{2x_\mathrm{cut}}+\frac{\ln V_\mathrm{cut}}{2x_\mathrm{cut}}+O(1) ,
\lb{WKB_CD_1st}
\eqn
where $j$ is a (large) integer, $V_\mathrm{cut}=V(x_\mathrm{cut})$, and we have expanded $\frac{k-\omega}{k+\omega}$ in terms of $\frac{1}{\omega}$.

Both the real and imaginary parts of the quasinormal frequencies bear a strong resemblance to Eq.~\eqref{DeltaCD_ratio}, obtained in the preceding subsection for a specific case.
This result confirms the observed tendency of the above numerical and semianalytical results in the literature~\cite{agr-qnm-21,agr-qnm-22} regarding the low-lying modes.
A certain degree of uncertainty comes from the term $\ln(D/C)$, which might affect the coefficient of the subleading contributions, as in the case of Eq.~\eqref{DeltaCD_ratio}.
For a general approach, unfortunately, a more detailed specification is out of reach.
Again, the resulting asymptotic behavior for the spectrum is found to be suppressed by the location of the cut.
Nonetheless, the ratio between the real and imaginary parts is independent of $x_\mathrm{cut}$ and is used to obtain the green dashed curve in Fig.~\ref{schematic_qnm}.

\subsection{C. General results for a potential with higher-order discontinuity}

As has been speculated and numerically shown in Ref.~\cite{agr-qnm-22}, the asymptotic properties of the quasinormal mode spectrum do not change if one replaces the step function by a continuous piecewise linear one.
Therefore, one might wonder whether the resultant quasinormal mode spectrum will be significantly modified if a more moderate discontinuity is introduced to the potential.
In this subsection, we show in a more general context that this is not the case.

We refine the potential introduced in the previous section further by replacing the ``step'' with an $n$th-order discontinuity in the potential $V(x)$.
For consistency in notation, the point of discontinuity will still be denoted by $x_\mathrm{cut}$, except now the potential is smoothly connected at the point up to $(n-1)$th order.
Also, we assume that the WKB approximation is valid on both sides.
Therefore, the wave function for $x < x_\mathrm{cut}$ still possesses the form of Eq.~\eqref{wf_WKB}, while for $x> x_\mathrm{cut}$ we similarly have
\bqn
g(\omega, x)=Ee^{iS(x_\mathrm{cut}, x)}+Fe^{-iS(x_\mathrm{cut}, x)} .
\lb{wg_WKB}
\eqn
where the coefficients $E, F$ are determined (reminiscent of $C, D$) to ensure the boundary condition at $x\to +\infty$ given in Eq.~\eqref{pt_boundary}.

By following a similar procedure to that which led to Eq.~\eqref{WKB_Wronskian_1st}, the resultant Wronskian evaluated at $x=x_\mathrm{cut}$ reads
\bqn
W(\omega)=(E+F)(iS'(x_\mathrm{cut}^-))(Ce^{iS_\mathrm{cut}}-De^{-iS_\mathrm{cut}})-(Ce^{iS_\mathrm{cut}}+De^{-iS_\mathrm{cut}})(iS'(x_\mathrm{cut}^+))(E-F)=0 ,\nonumber\\
\lb{WKB_Wronskian_nth}
\eqn
where $S'(x_\mathrm{cut}^-) = \left.S'(x)\right|_{x\to x_\mathrm{cut}^-}$ and $S'(x_\mathrm{cut}^+) = \left.S'(x)\right|_{x\to x_\mathrm{cut}^+}$.
The above equation can be reorganized to read
\bqn
\frac{\left(\frac{E+F}{E-F}-1\right)S'(x_\mathrm{cut}^-)+\left(S'(x_\mathrm{cut}^-) -S'(x_\mathrm{cut}^+)\right)}{\frac{E+F}{E-F}S'(x_\mathrm{cut}^-) +S'(x_\mathrm{cut}^+)}=\frac{D}{C}e^{-2iS_\mathrm{cut}} .
\lb{WKB_qnm_nth}
\eqn

Before proceeding further, we note that at the limit of our interest, namely, the quasinormal frequency with an arbitrarily large (negative) imaginary part, we have
\bqn
\lim\limits_{\mathrm{Im}\omega \to -\infty} \frac{E+F}{E-F}= 1 .
\lb{WKB_limit_FE}
\eqn
By observing Eq.~\eqref{wg_WKB}, in the limit of large frequency, one may ignore the contribution from the potential and approximately consider $k\sim \omega$.
Subsequently, the term associated with $F$ behaves asymptotically as $e^{-i\omega x}$.
As a result, for very large quasinormal frequencies $\mathrm{Im}\omega \to -\infty$, it blows up exponentially.
However, Eq.~\eqref{wg_WKB} is known to satisfy the outgoing wave boundary condition given in the second line of Eq.~\eqref{pt_boundary} at $x\to +\infty$.
Therefore, the only possibility is that $F \equiv F(\omega)$, when analytically continued to the complex plane, must vanish as $\mathrm{Im}\omega \to -\infty$, and thus Eq.~\eqref{WKB_limit_FE} is obtained.
To be more precise, not only does $\lim\limits_{\mathrm{Im}\omega \to -\infty} \frac{F}{E}= 0$, but it also approaches zero drastically fast in order to suppress another exponentially increasing factor.

An additional concern is that Eq.~\eqref{S_WKB} is no longer valid in the present context~\cite{WKB-02}; otherwise it can be shown that Eq.~\eqref{WKB_qnm_nth} leads to a contradiction.
This is because as the potential is a continuous function $V(x_\mathrm{cut}^-)=V(x_\mathrm{cut}^+) $, Eq.~\eqref{S_WKB} implies that $S'(x_\mathrm{cut}^-) =S'(x_\mathrm{cut}^+)$.
As a result, in the limit $\mathrm{Im}\omega \to -\infty$, Eq.~\eqref{WKB_limit_FE} indicates that the l.h.s. of Eq.~\eqref{WKB_qnm_nth} vanishes while the r.h.s. increases substantially.

The above difficulty can be resolved by considering a higher-order WKB approximation~\cite{WKB-02}.
Instead of Eq.~\eqref{S_WKB}, one has
\bqn
S(x_0,x) &=& \int_{x_0}^{x} dx' \left\{ k(x') +  \frac{ik'(x')}{2k(x')} + (i)^2\left[\frac{k''(x')}{(2k(x'))^2}-\frac{3k'(x')^2}{(2k(x'))^3}\right]+ \cdots + (i)^n\left[\frac{k^{(n)}(x')}{(2k(x'))^n}+\cdots\right]+\cdots\right\} \nonumber \\
&\simeq& \int_{x_0}^{x} dx' \left[ k(x') +  \frac{ik'(x')}{2k(x')} + \frac{(i)^2k''(x')}{(2k(x'))^2}+ \cdots + \frac{(i)^nk^{(n)}(x')}{(2k(x'))^n}+\cdots\right] .
\lb{S_WKB_nth}
\eqn
In deriving the above expression, for a given order of $\hbar$, only the terms which eventually lead to a discontinuity at $S'(x_\mathrm{cut})$ will be considered. 
Moreover, in the second line, among different combinations, one only retains the term where the discontinuity is originated from that in the derivatives of $V$ of the largest possible order. 
By observing Eq.~\eqref{S_WKB_nth}, it becomes apparent that the numerator on the l.h.s. of Eq.~\eqref{WKB_qnm_nth} indeed does not vanish, as it is dictated by the lowest-order discontinuity present in the potential.
Subsequently, the latter receives its dominant contribution from the discontinuity of the $n$th derivative of $k(x)=\sqrt{\omega^2-V(x)}$.

If one assumes that the discontinuity of the potential is at $n$th order, for large frequency, the l.h.s. of Eq.~\eqref{WKB_qnm_nth} gives 
\bqn
\frac{-i^n  }{(2\omega)^{n+2}}\Delta\left.\left(\frac{d^nV}{dx^n}\right)\right|_{x_\mathrm{cut}} ,\nonumber
\eqn
and after some algebra, the resultant asymptotic quasinormal normal modes are
\bqn
\omega_R &=& \frac{\pi j}{x_\mathrm{cut}} + O(1) ,\nonumber \\
\omega_I &=& -\frac{\ln(\pi j)}{x_\mathrm{cut}}+\frac{\ln x_\mathrm{cut}}{x_\mathrm{cut}} - \frac{\ln(D/C)}{2x_\mathrm{cut}}+\frac{\ln \Delta\left.\left(\frac{d^nV}{dx^n}\right)\right|_{x_\mathrm{cut}}}{2 x_\mathrm{cut}}+O(1) .
\lb{WKB_CD_nth}
\eqn
It is noted that the second term on the r.h.s. of Eq.~\eqref{S_WKB_nth}, when evaluated for $S_\mathrm{cut}$, gives rise to a contribution to the exponential on the r.h.s. of Eq.~\eqref{WKB_qnm_nth}, which is found to be proportional to $\ln k(x_\mathrm{cut})\sim \ln\omega$.
However, as its coefficient is smaller by a factor of $\hbar$ (not explicitly shown in the natural units adopted by us), it is not included in the resultant expression.
The consistency of Eq.~\eqref{WKB_CD_nth} can be readily ascertained as it becomes Eq.~\eqref{WKB_CD_1st} when taking the order of discontinuity $n=0$.

\section{IV. Relation with other conventional methods}

The asymptotic results obtained in the previous section are consistent with the numerical results presented in Sec.~II, as well as with the low-lying modes found for the piecewise potential in Refs.\cite{agr-qnm-21,agr-qnm-22}.
On the other hand, as mentioned above, it is drastically different from those obtained by various conventional approaches, namely, the WKB method~\cite{agr-qnm-WKB-01}, the continued fraction method~\cite{agr-qnm-continued-fraction-01,agr-qnm-continued-fraction-02,agr-qnm-continued-fraction-03}, and the monodromy method~\cite{agr-qnm-05}, among others.
In a rather general context, they all give rise to similar asymptotic behavior of quasinormal frequencies with a finite real part and large imaginary part with equal spacing.
In particular, recent studies~\cite{agr-qnm-Poschl-Teller-03,agr-qnm-Poschl-Teller-04} concerning a piecewise P\"oschl-Teller potential also indicate distinct results from ours.
In this regard, it is essential to clarify the difference between the present approach and other more conventional methods for quasinormal modes and discuss the related physical interpretations.

First, let us briefly review how the WKB method~\cite{agr-qnm-WKB-01} leads to the asymptotic quasinormal spectrum lining up along the imaginary axis.
For a quasinormal mode solution, the amplitudes of the transmitted and reflected waves are of the same order of magnitude. 
This is a direct consequence of the probability flux conservation as the boundary condition of quasinormal modes does not allow for an incoming wave. 
Moreover, in the case of the WKB approximation, the transmitted amplitude can be estimated by $e^{-B}$.
Here, $B$ is given by an integral of $\sqrt{V-\omega^2}$ between the two classical turning points of the potential, where $k(x)$ vanishes.
But since $\omega^2 < V(x)$ between the two turning points, it mostly leads to an exponentially small ratio of the transmitted amplitude to the reflected one.
Therefore, the condition for a quasinormal mode corresponds to a particular case, sometimes referred to as the second-order turning point.
For small overtone numbers, the latter implies that $\sqrt{V-\omega^2} \sim 0$, where the frequency $\omega$ can be complex.
The resultant quasinormal frequencies can be evaluated based only on the information of the effective potential (as well as its derivatives) at its maximum~\cite{agr-qnm-WKB-01,agr-qnm-WKB-03,agr-qnm-WKB-05}.
Consequently, at first glance, a ``cut'' planted further away from the maximum of the potential should not affect the above calculations.

However, to investigate the asymptotic properties of the quasinormal modes obtained when employing the WKB approximation, in principle, one has to deal with a contour integral in the complex coordinate space.
In particular, for the complex frequency with a large imaginary part, the relevant turning points (originally located on the real axis) will also migrate to the complex plane.
Therefore, caution must be taken in choosing an appropriate counter in accordance with the branching cuts~\cite{WKB-02,WKB-03}.
In particular, as shown schematically by Fig.~2 of Ref.~\cite{agr-qnm-WKB-07}, all of the branching cuts are located in the region with finite radius.
The counter $C_0$ can be subsequently deformed and divided into three parts, with one of them being an integration on an infinitely large circle.
The resulting behavior of the quasinormal mode spectrum is found to be largely consistent with those obtained by other methods, except that the real part of the frequency vanishes asymptotically.
The discontinuity introduced in our approach, however, leads to an additional branching out.
The latter emanates from $r=r(x_\mathrm{cut})$ and stretches to infinity.
As a result, the procedure carried out above, which leads to an estimation of the higher-overtone modes, cannot be applied straightforwardly to the present case.
Our analyses indicate that the above procedure cannot be applied straightforwardly due to the infeasibility of analytic continuation of the wave function around the ``cut'' on the real axis.
Equivalently, one may consider that there is a branching cut which passes through the point of discontinuity.
Therefore, the contributions from the contour integration on an infinitely large circle have to be replaced with those on one edge of the branching cut.
The latter turns out to be rather significant, which in turn, gives rise to a large impact on the asymptotic quasinormal mode spectrum.

Similarly, the monodromy method is also carried out in terms of the analytic continuation of the wave function in coordinate space.
In particular, as illustrated in Fig.~2 of Ref.~\cite{agr-qnm-05}, two branching cuts are involved.
One of them is related to the singularity, which can be shown to be irrelevant by avoiding it through an appropriate choice of the branching cut's orientation. 
The other branching cut, which originates from the horizon, is crucial to the physical problem.
It can be chosen conveniently to give rise to the ratio of the wave functions after the contour completes a counterclockwise circle around the horizon at $r=r_h=1$.
This approach was developed further by Andersson and Howls~\cite{agr-qnm-WKB-08} by employing the complex WKB method.
In terms of Stokes lines and constants, the form of the analysis becomes apparently simpler, and the same results regarding the highly damped quasinormal frequencies are obtained. 
When an additional discontinuity is added to the effective potential, again, one may imagine that a new branching cut emanating from $r=r(x_\mathrm{cut})$ is subsequently planted.
As a result, the analytic continuation cannot be performed straightforwardly. 
To be specific, the monodromy method relies on a deliberate choice of an enclosed contour, which avoids all possible branching cuts and singularities. 
One observes that both of the above approaches involve a part of the contour that loops around an infinitely large circle. 
However, the additional branching cut discussed above introduces novel complications which prohibit a straightforward application of the residue theorem. 
Moreover, the entire analysis of the monodromy method is based on the assumption that $\mathrm{Im}\omega \gg \mathrm{Re}\omega$, which is no longer valid in the present case.

The continued fraction method is known for its high precision and versatility in the study of quasinormal modes.
To be specific, the method relies on the expansions of the wave function and the effective potential around the horizon $r=1$.
However, the expansion of the effective potential becomes infeasible when a discontinuity is introduced. 
Subsequently, it is not surprising that the asymptotic results are also modified.

Moreover, it is essential to point out that the asymptotic quasinormal frequencies found above in the modified P\"oschl-Teller potential are very different from those obtained in Ref.~\cite{agr-qnm-Poschl-Teller-03,agr-qnm-Poschl-Teller-04}.
In Ref.~\cite{agr-qnm-Poschl-Teller-03}, the ``step'' was introduced precisely at $x_\mathrm{cut}= 0$.
Also, the derivation was based on the assumption that the quasinormal frequency possesses a large positive imaginary part but a moderate real part.
In the above derivation, however, the condition $x_\mathrm{cut}\gg 1$ plays a crucial role, and subsequently leads to different results.
On the other hand, by introducing a ``spike'' into the effective potential, the resultant quasinormal frequencies have been shown to be similar to our results~\cite{agr-qnm-29}.
In this regard, it seems that a discontinuity planted at the maximum of the effective potential leads to substantial differences.
This aspect certainly deserves further investigation.

One potentially promising approach is the venerable Chandrasekhar-Detweiler method~\cite{agr-qnm-07}, which does not seem to be affected by the discontinuity.
In order to avoid numerical instability, the master equation is first reformulated into a Riccati equation in terms of a phase function $\phi(x)$.
Subsequently, the quasinormal frequencies can be obtained by numerically integrating the resultant equation from both ends while requiring that the two solutions meet each other at an arbitrary point in the middle.
It is straightforward to show that such a connection condition corresponds to the vanishing of the Wronskian, in the case of the original master equation.
The numerical integration does not suffer from the instability, which would significantly undermine the validity of the procedure if it were directly applied to the original master equation.
In particular, Chandrasekhar and Detweiler pointed out in their paper that the procedure is capable of determining the quasinormal frequency so long as $|\mathrm{Re}\omega| \ge |\mathrm{Im}\omega|$.
As the quasinormal modes of the Regge-Wheeler potential usually climb up the imaginary axis, the method was only used to determine the first few modes.
For the present case, we note that the approach is not beset by the discontinuity. 
Moreover, the condition $|\mathrm{Re}\omega| \ge |\mathrm{Im}\omega|$ is rather favorable to us as the resultant quasinormal frequencies actually go along the real axis.

\section{V. Possible astrophysical implications}

Besides being a part of the piecewise function approximation to a realistic black hole metric, one might wonder if the discontinuity is physically meaningful.
We speculate that a discontinuity might naturally take place in realistic physical scenarios.
If this is the case, then the modification of the gravitational quasinormal modes discussed above might lead to direct implications of astrophysical relevance.
The modified asymptotic behavior due to the discontinuity might be observed as it affects the signal-to-noise ratio of space-borne laser interferemeters~\cite{agr-SNR-05,agr-SNR-06}. 
Besides gravitational-wave detection, the black hole shadow is also one of the promising observables for the strong-field regime, as it is rather sensitive to the details of the matter distribution and the resultant effect might be experimentally relevant.
Its connection with the quasinormal modes~\cite{agr-shadow-09}, especially regarding the discontinuity, is also worth exploring.

In practice, a discontinuity in matter distribution may occur due to the surface of the compact star, cuspy halo, or related to an unidentified form of matter and energy.
In what follows, we discuss two explicit examples that are potentially meaningful in the context of astrophysics.

First, let us consider a simple scenario where a thin mass shell $\delta M$ is wrapped around a Schwarzschild black hole metric at the radius $r=r_{\mathrm{shell}}$, located beyond the innermost stable circular orbit $r_{\mathrm{shell}}> r_{\mathrm{ISCO}}=3r_h$.
In this case, the effective potential reads
\bqn
V_{\mathrm{step}}=
\left\{\begin{array}{cc}
V_\mathrm{RW}(r_h, r)     &  r\le r_{\mathrm{shell}}  \cr\\
V_\mathrm{RW}(r_h', r)     &  r> r_{\mathrm{shell}}  
\end{array}\right. ,
\lb{V_step_thin_shell}
\eqn
where $r_h=2M, r_h'=2(M+\delta M)$.
This subsequently gives rise to a ``step up'' in the effective potential for scalar perturbations, and a ``step down'' for the vector-type gravitational perturbations.
For both cases, the findings of the present study can be readily applied.

Second, for a spherically symmetric compact object, the matter distribution is governed by the Tolman-Oppenheimer-Volkoff equation~\cite{book-general-relativity-Carroll}
\bqn
\frac{dp}{dr}=-\frac{(\rho+p)[m(r)+4\pi  r^3p]}{r[r-2m(r)]} .
\lb{TOV}
\eqn
In this case, a discontinuity may appear on the surface of the star, an interior interface separating two distinct layers of matter, or a sharp edge of an accretion disk.
The last case may occur, for instance, due to orbital instability~\cite{agr-accretion-disk-05,agr-accretion-disk-07} such as when the inner edge of the accretion disk coincides with the innermost stable circular orbit.
In general, the difference may be characterized by distinct equations of state, namely, $\rho_1(p)\ne\rho_2(p)$, and a discontinuity is present in the first-order derivative of the pressure.
This subsequently leads to a discontinuity in the energy-momentum tensor and therefore the backreaction received by the spacetime metric.
Our derivation indicates that no matter how insignificant such a discontinuity is, the asymptotic behavior of the quasinormal modes will be significantly modified.
Indeed, one family of standard $w$ modes, known as the curvature modes, is observed to possess similar asymptotic behavior to that found in the piecewise potential~\cite{agr-qnm-27}.
To be specific, the imaginary part of the quasinormal frequencies increases moderately with the overtone number, while their real part increases more rapidly.

\section{VI. Further discussions and concluding remarks}

In this work, we investigated the resultant modification in the quasinormal mode spectrum due to the piecewise approximate potential.
The study was motivated by recent findings on the apparent contradiction between the evolution of perturbations in the time domain and quasinormal mode spectrum in the frequency domain.
While the temporal evolution gives rise to a desirable match with the case of the Regge-Wheeler potential, the resultant asymptotic quasinormal modes present a distinct feature.
The spectrum is found to mostly lie along the real axis, which appears rather different from that obtained by using the realistic potential for physical black hole metrics.
Subsequently, the present study thus involved an attempt to answer the points raised at the beginning of the paper.
Our investigation consisted of numerical as well as analytical approaches.

As for the numerical approach, we devised a modified Regge-Wheeler potential in order to separate the two features related to the fundamental mode and the asymptotic behavior of the spectrum.
The asymptotic quasinormal modes are thus attributed to the discontinuity introduced by the piecewise function approximation.
In particular, our findings remain valid even though the strength of the ``step'' is insignificant and/or located further away from the black hole horizon.
The analytic derivation was initiated by exploring a modified P\"oschl-Teller potential.
The analytic arguments were then generalized.
The discussions were first extended to the case of an arbitrary form of the potential where the WKB approximation is valid on both sides of the discontinuity.
Then, we explored the case where the discontinuity in the potential originates from a higher-order derivation.
As a result, we demonstrated that the numerically observed feature is sound in a rather general context.

The derivation given in this paper was compared against other standard methods.
For instance, the WKB method evaluates the quasinormal frequencies by using merely the derivatives of the effective potential at its maximum.
Also, by replacing the Regge-Wheeler potential by a P\"oschl-Teller one mainly around its maximum, a reasonable agreement can be achieved.
The above well-known results seem to indicate that only the region near the black hole is relevant for the quasinormal modes.
Also, the monodromy method is usually employed to handle the asymptotic properties of the quasinormal modes with a large imaginary part.
When compared with the case of a piecewise potential, the observed distinction can be traced back to the infeasibility of analytic continuation or the additional branching cut caused by the discontinuity.
The results discussed in the present study were examined in a consistent framework with other conventional methods and related physical interpretations.
It was argued that some specific methods cease to apply to this particular case since either an analytic continuation becomes nontrivial or the Taylor expansion turns out to be inaccessible due to the discontinuity.
Some subtleties regarding the details of the technique were clarified and apprehended.

The Prony method utilized in the present study has been shown to be an efficient tool for evaluating the low-lying quasinormal modes.
In conjunction with appropriately tuned numerical integration for the temporal evolution, it can successfully extract the first few overtones of the Regge-Wheeler potential.
The method is rather universal as it does not depend on the specific form of the potential, and the fitting process is quite efficient.
The downside of the approach is that it cannot be used for quasinormal modes of higher overtone.
Moreover, one cannot straightforwardly estimate the error bound of the extracted frequencies.
In certain circumstances, the subtracted frequencies are found to be sensitive to the range where the fit of the Prony method is carried out.
As a result, we had to resort to an analytic approach for the asymptotic behavior of the quasinormal modes.
Therefore, it is desirable to also evaluate the quasinormal frequencies by using other independent approaches, such as the Chandrasekhar-Detweiler method.

There are, however, still a few issues that deserve to be explored further.
The difference observed in the quasinormal mode spectrum between the piecewise potential and the Regge-Wheeler one was only partly resolved.
We found that as the quasinormal modes lie along the real axis, their imaginary parts are of the same order of magnitude.
As a result, such modes may constitute the resultant waveform with similar importance.
In other words, the observed quasinormal oscillations cannot be overwhelmed by one single mode.
On the other hand, when the quasinormal modes line up along the imaginary axis, the fundamental mode has a more significant role than others.
The above argument partly explains why the fundamental modes in the two cases do not match each other, even though the long-term temporal evolutions are identical.
However, as pointed out by other authors~\cite{agr-qnm-21,agr-qnm-22}, a deeper understanding of the problem stems from the completeness of quasinormal modes~\cite{agr-qnm-28,agr-qnm-Poschl-Teller-05,agr-qnm-19,agr-qnm-20,agr-qnm-29}.
Indeed, this is a rather intriguing topic.
If one can show that the quasinormal modes are complete, any physical solution is then expected to be represented as a sum of quasinormal modes.
Less ambitiously~\cite{agr-qnm-21}, even if the quasinormal modes do not form a complete set, it would still be rewarding if one manages to capture the main characteristics of the physical system quantitatively using a few low-lying fundamental modes~\cite{agr-qnm-29}.
Besides, it is meaningful to understand whether the information on quasinormal modes can be unambiguously extracted from the time evolution even though other components persist.
Moreover, discontinuities are physically relevant, as they could be present in matter distribution associated with space dust or the surface of a compact object.
Therefore, it might be interesting to study implications regarding possible observations owing to the different quasinormal modes.

We pointed out that there is a ``paradox'' regarding the resultant quasinormal mode spectrum.
To be specific, it seems to be indistinguishable from a numerical viewpoint whether the computation is carried out for the Regge-Wheeler potential or a piecewise one.
This is because the algorithm is not aware of any sudden jump where the potential has not been sampled. 
The above consideration seems to indicate a substantial uncertainty about the numerical outcome.
Nonetheless, we speculate that there might be one possible scenario to resolve the seeming ``paradox", i.e., when one utilizes a piecewise approximate potential with sufficiently high resolution.
The resultant low-lying modes from the numerical calculations would be largely identical to those of the black hole metric, while the asymptotic spectrum would still follow those of the piecewise potential.
We note that such an assertion remains to be confirmed by straightforward calculations, and it would be interesting to systematically investigate the uncertainty associated with the precision of the sampling process.

Recently, the effects of perturbative but continuously deformed effect potential have been investigated~\cite{agr-qnm-26}.
There, the impact on the quasinormal modes was studied with respect to different underlying theories of gravity.
From our viewpoint, we argued that discontinuity plays a physically relevant role, and focused on asymptotic properties of the resultant quasinormal mode spectrum.
Further studies along this direction are in progress.

\section*{Acknowledgments}
We wish to thank Yunqi Liu, Peng Liu, Hang Liu, and Song-Bai Chen for enlightening discussions.
We gratefully acknowledge the financial support from
Funda\c{c}\~ao de Amparo \`a Pesquisa do Estado de S\~ao Paulo (FAPESP),
Funda\c{c}\~ao de Amparo \`a Pesquisa do Estado do Rio de Janeiro (FAPERJ),
Conselho Nacional de Desenvolvimento Cient\'{\i}fico e Tecnol\'ogico (CNPq),
Coordena\c{c}\~ao de Aperfei\c{c}oamento de Pessoal de N\'ivel Superior (CAPES),
and National Natural Science Foundation of China (NNSFC) under contract Nos. 11805166, 11775036, and 11675139.
A part of this work was developed under the project Institutos Nacionais de Ciências e Tecnologia - Física Nuclear e Aplicações (INCT/FNA) Proc. No. 464898/2014-5.
This research is also supported by the Center for Scientific Computing (NCC/GridUNESP) of the S\~ao Paulo State University (UNESP).

\section{Appendix}
\renewcommand{\theequation}{A.\arabic{equation}}
\setcounter{equation}{0}

In this appendix, we derive the specific expressions that are utilized in the main text regarding the problem of the (modified) P\"oschl-Teller potential.
The formal solutions of the Schr\"odinger equation Eq.~\eqref{pt_homo_eq} with the potential $U=-{V}_\mathrm{PT}$ can be found in standard textbooks such as Ref.~\cite{book-quantum-mechanics-Flugge}.
One may conveniently choose two independent solutions with even and odd parities, namely, $u_e(-x)=u_e(x)$ and $u_o(-x)=-u_o(x)$, which read
\bqn
u_e(x) &=&\cosh^\lambda \kappa x {_2F_1}(a,b,\frac12;-\sinh^2 \kappa x) ,\nonumber\\
u_o(x) &=&\cosh^\lambda \kappa x \sinh \kappa x {_2F_1}(a+\frac12,b+\frac12,\frac32;-\sinh^2 \kappa x) ,
\lb{ueuo_solution}
\eqn
where
\bqn
a = \frac12\left(\lambda+i\frac{\omega}{\kappa}\right) ,\nonumber\\
b = \frac12\left(\lambda-i\frac{\omega}{\kappa}\right) ,
\lb{def_ab}
\eqn
and
\bqn
\lambda =  \frac12 + \frac{\sqrt{4V_m+\kappa^2}}{2\kappa} .
\lb{def_lambda}
\eqn
It is noted that $\lambda > 1$ when $V_m$ and $\kappa$ are positive real numbers.

We are aiming at an appropriate combination defined in Eq.~\eqref{f_mPT}, which agrees with the boundary condition given in the first line of Eq.~\eqref{pt_boundary}.
By using the expansion formulae of ${_2F_1}(a,b,\frac12;z)$ and ${_2F_1}(a,b,\frac32;z)$ at $z\to \infty$ to the first order, one has
\bqn
u_e(x) &\to& \Gamma\left(\frac12\right)\left\{\frac{\Gamma\left(\frac{-i\omega}{\kappa}\right)e^{i\frac{\omega}{\kappa}\ln 2}}{\Gamma\left(\frac{\lambda}{2}-i\frac{\omega}{2\kappa}\right)\Gamma\left(\frac{1-\lambda}{2}-i\frac{\omega}{2\kappa}\right)}e^{i\omega x}+
\frac{\Gamma\left(\frac{i\omega}{\kappa}\right)e^{-i\frac{\omega}{\kappa}\ln 2}}{\Gamma\left(\frac{\lambda}{2}+i\frac{\omega}{2\kappa}\right)\Gamma\left(\frac{1-\lambda}{2}+i\frac{\omega}{2\kappa}\right)}e^{-i\omega x}\right\} ,\nonumber\\
u_o(x) &\to& -\Gamma\left(\frac32\right)\left\{\frac{\Gamma\left(\frac{-i\omega}{\kappa}\right)e^{i\frac{\omega}{\kappa}\ln 2}}{\Gamma\left(\frac{\lambda+1}{2}-i\frac{\omega}{2\kappa}\right)\Gamma\left(\frac{2-\lambda}{2}-i\frac{\omega}{2\kappa}\right)}e^{i\omega x}+
\frac{\Gamma\left(\frac{i\omega}{\kappa}\right)e^{-i\frac{\omega}{\kappa}\ln 2}}{\Gamma\left(\frac{\lambda+1}{2}+i\frac{\omega}{2\kappa}\right)\Gamma\left(\frac{2-\lambda}{2}+i\frac{\omega}{2\kappa}\right)}e^{-i\omega x}\right\}  ,\nonumber\\
\lb{ueuo_asymptotic_1st_negativeX}
\eqn
for $x\to -\infty$, and 
\bqn
u_e(x) &\to& \Gamma\left(\frac12\right)\left\{\frac{\Gamma\left(\frac{-i\omega}{\kappa}\right)e^{i\frac{\omega}{\kappa}\ln 2}}{\Gamma\left(\frac{\lambda}{2}-i\frac{\omega}{2\kappa}\right)\Gamma\left(\frac{1-\lambda}{2}-i\frac{\omega}{2\kappa}\right)}e^{-i\omega x}+
\frac{\Gamma\left(\frac{i\omega}{\kappa}\right)e^{-i\frac{\omega}{\kappa}\ln 2}}{\Gamma\left(\frac{\lambda}{2}+i\frac{\omega}{2\kappa}\right)\Gamma\left(\frac{1-\lambda}{2}+i\frac{\omega}{2\kappa}\right)}e^{i\omega x}\right\} ,\nonumber\\
u_o(x) &\to& +\Gamma\left(\frac32\right)\left\{\frac{\Gamma\left(\frac{-i\omega}{\kappa}\right)e^{i\frac{\omega}{\kappa}\ln 2}}{\Gamma\left(\frac{\lambda+1}{2}-i\frac{\omega}{2\kappa}\right)\Gamma\left(\frac{2-\lambda}{2}-i\frac{\omega}{2\kappa}\right)}e^{-i\omega x}+
\frac{\Gamma\left(\frac{i\omega}{\kappa}\right)e^{-i\frac{\omega}{\kappa}\ln 2}}{\Gamma\left(\frac{\lambda+1}{2}+i\frac{\omega}{2\kappa}\right)\Gamma\left(\frac{2-\lambda}{2}+i\frac{\omega}{2\kappa}\right)}e^{i\omega x}\right\}  ,\nonumber\\
\lb{ueuo_asymptotic_1st_positiveX}
\eqn
for $x\to +\infty$.

Based on Refs.~\cite{agr-qnm-Poschl-Teller-01,agr-qnm-Poschl-Teller-02}, one considers the bound state where $\omega$ is real.
We introduce the transformation
\bqn
\omega\to \omega'=\omega(\kappa')
\lb{trans_PT_omega}
\eqn
together with those defined in Eq.~\eqref{trans_PT}, namely,
\bqn
\left\{\begin{array}{c}
x\to -ix'  \cr\\
\kappa \to i\kappa'  
\end{array}\right. . \nonumber
\eqn

After implementing the above substitution, $\kappa' x'$ continues to be real numbers so that the limits for the quantities such as $\kappa x$ and $z\equiv -\sinh^2\kappa x$ remain unchanged.
It is noted that the asymptotic forms for the wave functions transform from a bound state to out-going waves.
A tricky factor is that now $\lambda$ is complex owing to Eq.~\eqref{def_lambda}, which involves the substitution of $\kappa$.
Fortunately, one still has the asymptotic relation $\cosh^\lambda \kappa x(-z)^{-a}\to e^{i\frac{\omega}{\kappa}\ln 2}e^{-i\omega x}$, as it is easy to verify that the real part of $\lambda$ remains positive.

Subsequently, it is straightforward to find that
\bqn
A &=& {\Gamma\left(\frac{\lambda}{2}-i\frac{\omega}{2\kappa}\right)\Gamma\left(\frac{1-\lambda}{2}-i\frac{\omega}{2\kappa}\right)}, \nonumber \\
B &=& 2{\Gamma\left(\frac{\lambda+1}{2}-i\frac{\omega}{2\kappa}\right)\Gamma\left(\frac{2-\lambda}{2}-i\frac{\omega}{2\kappa}\right)} ,
\lb{AB_ueuo}
\eqn
from which one also encounters the specific forms for $C, D$ given in Eq.~\eqref{f_mPT_CD} by comparing against Eqs.~\eqref{ueuo_asymptotic_1st_positiveX},
\bqn
C &=& \Gamma\left(\frac12\right)\left\{
\frac{\Gamma\left(\frac{\lambda}{2}-i\frac{\omega}{2\kappa}\right)\Gamma\left(\frac{1-\lambda}{2}-i\frac{\omega}{2\kappa}\right)}
{\Gamma\left(\frac{\lambda}{2}+i\frac{\omega}{2\kappa}\right)\Gamma\left(\frac{1-\lambda}{2}+i\frac{\omega}{2\kappa}\right)}
+
\frac{\Gamma\left(\frac{\lambda+1}{2}-i\frac{\omega}{2\kappa}\right)\Gamma\left(\frac{2-\lambda}{2}-i\frac{\omega}{2\kappa}\right)}
{\Gamma\left(\frac{\lambda+1}{2}+i\frac{\omega}{2\kappa}\right)\Gamma\left(\frac{2-\lambda}{2}+i\frac{\omega}{2\kappa}\right)}
\right\} \Gamma\left(\frac{i\omega}{\kappa}\right)e^{i\frac{\omega}{\kappa}\ln 2}, \nonumber \\
D &=& 2\Gamma\left(\frac12\right)\Gamma\left(\frac{-i\omega}{\kappa}\right)e^{i\frac{\omega}{\kappa}\ln 2} .
\lb{CD_ueuo}
\eqn

For the reasons given in the main text, one also has to work out the expansions up to the second order in the limit $x\to +\infty$.
After some algebra, one finds
\bqn
u_e(x) &\to& \Gamma\left(\frac12\right)\left\{\left[1-\frac{\lambda(\lambda-1)}{\left(1+i\frac{\omega}{\kappa}\right)} e^{-2\kappa x}\right]\frac{\Gamma\left(\frac{-i\omega}{\kappa}\right)e^{i\frac{\omega}{\kappa}\ln 2}}{\Gamma\left(\frac{\lambda}{2}-i\frac{\omega}{2\kappa}\right)\Gamma\left(\frac{1-\lambda}{2}-i\frac{\omega}{2\kappa}\right)}e^{-i\omega x}\right. \nonumber\\
                                                             &+&\left.\left[1-\frac{\lambda(\lambda-1)}{\left(1-i\frac{\omega}{\kappa}\right)} e^{-2\kappa x}\right]\frac{\Gamma\left(\frac{i\omega}{\kappa}\right)e^{-i\frac{\omega}{\kappa}\ln 2}}{\Gamma\left(\frac{\lambda}{2}+i\frac{\omega}{2\kappa}\right)\Gamma\left(\frac{1-\lambda}{2}+i\frac{\omega}{2\kappa}\right)}e^{i\omega x}\right\} ,\nonumber\\
u_o(x) &\to& +\Gamma\left(\frac32\right)\left\{\left[1- \frac{\lambda(\lambda-1)}{\left(1+i\frac{\omega}{\kappa}\right)}e^{-2\kappa x}\right]\frac{\Gamma\left(\frac{-i\omega}{\kappa}\right)e^{i\frac{\omega}{\kappa}\ln 2}}{\Gamma\left(\frac{\lambda+1}{2}-i\frac{\omega}{2\kappa}\right)\Gamma\left(\frac{2-\lambda}{2}-i\frac{\omega}{2\kappa}\right)}e^{-i\omega x}\right.\nonumber\\
                                                               &+&\left.\left[1-\frac{\lambda(\lambda-1)}{\left(1-i\frac{\omega}{\kappa}\right)} e^{-2\kappa x}\right]\frac{\Gamma\left(\frac{i\omega}{\kappa}\right)e^{-i\frac{\omega}{\kappa}\ln 2}}{\Gamma\left(\frac{\lambda+1}{2}+i\frac{\omega}{2\kappa}\right)\Gamma\left(\frac{2-\lambda}{2}+i\frac{\omega}{2\kappa}\right)}e^{i\omega x}\right\}  .\nonumber\\
\lb{ueuo_asymptotic_2nd_positiveX}
\eqn
Subsequently, in the place of Eq.~\eqref{CD_ueuo}, we now have
\bqn
\tilde{C} &\equiv& C+\Delta C=C + \left(\Delta \tilde{C}_1 + \Delta \tilde{C}_2 \right)e^{-2\kappa x}, \nonumber \\
\tilde{D} &\equiv& D+\Delta D=D + \Delta \tilde{D} e^{-2\kappa x} ,\nonumber \\
\lb{CD_tilde_ueuo}
\eqn
where
\bqn
{\Delta \tilde{C}_1} &=& -\frac{\lambda(\lambda-1)}{\left(1-i\frac{\omega}{\kappa}\right)}
\frac{\Gamma\left(\frac{\lambda}{2}-i\frac{\omega}{2\kappa}\right)\Gamma\left(\frac{1-\lambda}{2}-i\frac{\omega}{2\kappa}\right)}
{\Gamma\left(\frac{\lambda}{2}+i\frac{\omega}{2\kappa}\right)\Gamma\left(\frac{1-\lambda}{2}+i\frac{\omega}{2\kappa}\right)} 
\Gamma\left(\frac12\right)\Gamma\left(\frac{i\omega}{\kappa}\right)e^{i\frac{\omega}{\kappa}\ln 2} ,\nonumber\\
{\Delta \tilde{C}_2}&=& -\frac{\lambda(\lambda-1)}{\left(1-i\frac{\omega}{\kappa}\right)}
\frac{\Gamma\left(\frac{\lambda+1}{2}-i\frac{\omega}{2\kappa}\right)\Gamma\left(\frac{2-\lambda}{2}-i\frac{\omega}{2\kappa}\right)}
{\Gamma\left(\frac{\lambda+1}{2}+i\frac{\omega}{2\kappa}\right)\Gamma\left(\frac{2-\lambda}{2}+i\frac{\omega}{2\kappa}\right)}
\Gamma\left(\frac12\right)\Gamma\left(\frac{i\omega}{\kappa}\right)e^{i\frac{\omega}{\kappa}\ln 2} , \nonumber \\
{\Delta \tilde{D}} &=& -\frac{2\lambda(\lambda-1)}{\left(1+i\frac{\omega}{\kappa}\right)}
\Gamma\left(\frac12\right)\Gamma\left(\frac{-i\omega}{\kappa}\right)e^{i\frac{\omega}{\kappa}\ln 2} .
\lb{CD_tilde_delta_ueuo}
\eqn
It is readily verfied $\Delta \tilde{C}_1, \Delta \tilde{C}_2 \ll C$ and $\Delta \tilde{D} \ll D$ in the limit of large frequency $\omega$.
Also, as is utilized in the main text, we have the following ratio\footnote{In the limit where $\Re\omega \gg 1$,
the argumet of the relevant $\Gamma$ function approaches the imaginary axis where the behavior of the function is modest.
We also note since $\lambda$ is complex, the pairs shown in the numerator and denominator are not precisely complex conjugates $\Gamma(z)$ and $\Gamma(\bar{z})$.
As a result, the approximation given in Eq.~\eqref{DeltaCD_ratio} does not hold when the real part of $\Re\omega \sim O(1)$, and in particular, when $\Im\omega \gg 1$.
For the latter case, the ratio on the l.h.s. of Eq.~\eqref{DeltaCD_ratio} possesses poles associated with the quasinormal modes of the original black hole.}:
\bqn
\frac{D}{\Delta \tilde{C}_1+\Delta \tilde{C}_2} \sim \left[\frac{i\omega}{\kappa\lambda(\lambda-1)}-\frac{1}{\lambda(\lambda-1)}\right] e^{i\phi} .
\lb{DeltaCD_ratio}
\eqn
In the derivation, one makes use of the properties $\overline{\Gamma(z)}=\Gamma(\bar{z})$ and notices that $\lambda$ is a complete but finite number.
We note that $\phi$ is a phase which will give a further correction to the first line of Eq.~\eqref{qnm_mPT}, but the resultant expression found in the text remains valid.

\bibliographystyle{h-physrev}
\bibliography{references_qian}

\end{document}